\documentclass{PoS}
\usepackage{graphicx}
\usepackage{amsfonts}
\usepackage{amssymb}
\usepackage{amsmath}

\title{
Dirac spectrum representation of Polyakov loop fluctuations in lattice QCD
\thanks{
T. M. D. is supported by Grant-in-Aid for 
JSPS Fellows (Grant No. 15J02108), and H. S. is supported 
by the Grants-in-Aid for Scientific Research (Grant 
No. 15K05076) from Japan Society for the Promotion of Science. 
K. R. acknowledges partial support of the U.S. 
Department of Energy under Grant No. DE-FG02- 
05ER41367, and fruitful discussions with Bengt Friman 
and Pok Man Lo. 
The work of K. R. and C. S. has been partly 
supported by the Polish Science Foundation (NCN) under 
Maestro Grant No. DEC-2013/10/A/ST2/00106 and by the 
Hessian LOEWE initiative through the Helmholtz 
International Center for FAIR. The lattice QCD calcula- 
tions were performed on NEC-SX8R and NEC-SX9 at 
Osaka University.}
}

\ShortTitle{Dirac spectrum representation of Polyakov loop fluctuations in lattice QCD}

\author{\speaker{Takahiro M. Doi}\\
        Department of Physics, Kyoto University, Kyoto 606-8502, Japan\\
        E-mail: \email{doi@ruby.scphys.kyoto-u.ac.jp}}

\author{Krzysztof Redlich\\
        Institute of Theoretical Physics, University of Wroclaw, PL-50204 Wroclaw, Poland\\
        Extreme Matter Institute EMMI, GSI, Planckstr. 1, D-64291 Darmstadt, Germany\\
        Department of Physics, Duke University, Durham, North Carolina 27708, USA
}

\author{Chihiro Sasaki\\
        Institute of Theoretical Physics, University of Wroclaw, PL-50204 Wroclaw, Poland\\
        Frankfurt Institute for Advanced Studies, D-60438 Frankfurt am Main, Germany
}

\author{Hideo Suganuma\\
        Department of Physics, Kyoto University, Kyoto 606-8502, Japan
}

\abstract{
Dirac spectrum representations of the Polyakov loop fluctuations 
are derived on the temporally odd-number lattice, where the temporal length is odd 
with the periodic boundary condition. 
We investigate the Polyakov loop fluctuations 
based on these analytical relations. 
It is semi-analytically and numerically found that 
the low-lying Dirac eigenmodes have little contribution to the Polyakov loop fluctuations, 
which are sensitive probe for the quark deconfinement. 
Our results suggest no direct one-to-one corresponding between 
quark confinement and chiral symmetry breaking in QCD. 
}

\FullConference{The 33rd International Symposium on Lattice Field Theory\\
		14 -18 July 2015\\
		Kobe International Conference Center, Kobe, Japan*}

\begin{document}

\def\slash#1{\not\!#1}
\def\slashb#1{\not\!\!#1}
\def\slashbb#1{\not\!\!\!#1}

\section{Introduction}
It is one of the most important problems in particle and nuclear physics 
to understand the nonperturbative properties of QCD, 
such as confinement and chiral symmetry breaking, including these relation. 
By numerous efforts, 
these properties have been partially understood. 

The Polyakov loop is an order parameter of quark confinement in the quenched QCD \cite{Rothe}. 
Moreover, recently it is pointed out that 
the Polyakov loop fluctuations are very sensitive probes for the quark deconfinement 
even in full QCD with light quarks \cite{LFKRS13}. 

The low-lying eigenmodes of the Dirac operator 
are important for chiral symmetry breaking, 
for example known as Banks-Casher relation \cite{BanksCasher}. 
At low temperature, the low-lying Dirac modes exist and have dominant contribution to chiral condensate 
while at high temperature these modes disappear and chiral symmetry is restored. 

In the presence of light dynamical quarks, 
both quark deconfinement and chiral restoration are not phase transition but crossover 
and take place in the same temperature region \cite{Karsch, ColumbiaPlot, aoki, O(4)}. 
This observation seems evidence 
that confinement and chiral symmetry breaking are strongly correlated in QCD. 
However, there are some contrary observations 
that chiral restoration does not immediately mean the quark deconfinement. 
For example, ``hadrons" can be observed as bound states 
after removal of low-lying Dirac modes \cite{LS11}. 
Also, it is shown that low-lying Dirac modes are not important for confinement properties 
such as the Polyakov loop and the linear confining  potential of a quark-antiquark system \cite{GIS}. 

In this paper, we derive some analytical relations to express 
the Polyakov loop and its fluctuations by Dirac eigenmodes. 
Then we semi-analytically and numerically show 
that low-lying Dirac modes have negligible contribution to the Polyakov loop fluctuations 
based on these analytical relations. 
This talk is mainly based on our recent work \cite{DRSS}. 
\section{Polyakov loop fluctuations and Dirac mode on a lattice}
In this section, we review the Polyakov loop fluctuations \cite{LFKRS13} 
and Dirac-mode expansion \cite{GIS} 
as the basics of our work. 
In this paper, 
we consider the ${\rm SU}(N_{\rm c})$ lattice QCD formalism on a square lattice with spacing $a$. 
We denote each site as $s=(s_1, s_2, s_3, s_4)$ with $s_\mu=1,2,\cdots,N_\mu$ 
and a link variable as $U_\mu(s)={\rm e}^{iagA_\mu(s)}$ with a gauge field, $A_\mu(s) \in su(N_c)$, 
and the gauge coupling $g$. 
We use spatially symmetric lattice, i.e., $N_1=N_2=N_3\equiv N_\sigma$ and 
$N_4\equiv N_\tau$, with $N_\sigma \ge N_\tau$. 
We define all 
the 	$\gamma$-matrices to be hermite as $\gamma_\mu^\dagger=\gamma_\mu$. 
\subsection{Polyakov loop fluctuations}
For each gauge configuration, 
the Polyakov loop $L$ and the $Z_3$ rotated Polyakov loop $\tilde{L}$ are defined as 
\begin{align}
L \equiv
\frac{1}{N_{\rm c}V} \sum_s {\rm tr}_c
\{\prod_{i=0}^{N_\tau-1}U_4(s+i\hat{4})\}, 
\ \ \ \ \ \ \ \ 
\tilde{L}
\equiv 
L \mathrm{e}^{2\pi k i/3} \ \ \ (k=0,\pm1), \label{Polyakov}
\end{align}
where $\hat{\mu}$ is the unit vector in direction of $\mu$ in the lattice unit, 
$V=N_\sigma^3N_\tau$ is the volume of 4-dimensional lattice, 
$k=0$ is taken in the confinement phase, 
and $k$ in the deconfinement phase is chosen such that 
the transformed Polyakov loop $\tilde{L}$ lies in its real sector. 
Using the $Z_3$ rotated Polyakov loop $\tilde{L}$, 
three Polyakov loop susceptibilities are introduced as 
\begin{align}
&T^3\chi_A=\frac{N_\sigma^3}{N_\tau^3}[\langle |L|^2 \rangle-\langle |L| \rangle^2], \label{ChiA}\\
&T^3\chi_L=\frac{N_\sigma^3}{N_\tau^3}[\langle (L_L)^2 \rangle-\langle L_L \rangle^2], \label{ChiL}\\
&T^3\chi_T=\frac{N_\sigma^3}{N_\tau^3}[\langle (L_T)^2 \rangle-\langle L_T \rangle^2], \label{ChiT}
\end{align}
where $L_L \equiv {\rm Re}(\tilde{L})$ and $L_T \equiv {\rm Im}(\tilde{L})$,
and $\langle x \rangle$ denotes an average over all gauge configurations. 
Also, their ratios are introduced as, 
\begin{align}
R_A \equiv \frac{\chi_A}{\chi_L}, \ \ \ \ \ \ \ \ 
R_T \equiv \frac{\chi_T}{\chi_L}. \label{RART}
\end{align}
By finite-temperature lattice QCD calculation, 
it is found that the ratio $R_A$ is a very sensitive probe for the quark deconfinement 
even in the presence of light dynamical quarks \cite{LFKRS13}. 

\subsection{Operator formalism and Dirac mode on a lattice}
The link-variable operator $\hat{U}_{\pm\mu}$ is defined by the matrix element
\begin{align}
\langle s | \hat{U}_{\pm\mu} |s' \rangle=U_{\pm\mu}(s)\delta_{s\pm\hat{\mu},s'}. \label{LinkOp}
\end{align}
At the temporal boundary, we impose the temporal antiperiodicity to the link-variable operator as
\begin{eqnarray}
\langle {\bf s}, N_t|\hat U_4| {\bf s}, 1 \rangle 
=-U_4({\bf s}, N_t) 
\label{eq:LVthermal}
\end{eqnarray}
to impose the temporal antiperiodicity to the Dirac operator 
for finite-temperature formalism. 
Using the link-variable operator, 
the Polyakov loop is expressed as 
\begin{align}
L
=-\frac{1}{N_{\rm c}V}{\rm Tr}_c \{\hat U_4^{N_\tau}\} 
=\frac{1}{N_c V}
\sum_s {\rm tr}_c \{\prod_{n=0}^{N_t-1} U_4(s+n\hat t)\},
\label{PolyakovOp}
\end{align}
where Tr$_c$ denotes the functional trace,
${\rm Tr}_c \equiv \sum_s {\rm tr}_c$,
and ${\rm tr}_c$ is taken over color index.

Unlike the Polyakov loop, 
a functional trace of a product of the link-variable operators 
corresponding to the non-closed path vanishes:
\begin{align}
{\rm Tr}_c(\hat{U}_{\mu_1}\hat{U}_{\mu_2}\cdots\hat{U}_{\mu_{N_P}})
={\rm tr}_c\sum_s U_{\mu_1}(s)\cdots U_{\mu_{N_P}}(s+\sum_{k=1}^{N_P-1}\hat{\mu}_k)
\langle s+\sum_{k=1}^{N_P}\hat{\mu}_k|s\rangle
=0,
\label{nonclosed}
\end{align}
with $\sum_{k=1}^{N_P}\hat{\mu}_k\neq 0$ for any non-closed path
with its length $N_P$.
This notable property of the link-variable operator is satisfied due to 
the definition of the link-variable operator 
and easily understood by Elitzur’s theorem \cite{Elitzur} 
that a vacuum expectation value of a gauge-variant operator is zero. 

The Dirac operator $\hat{\slashb{D}}$ on the lattice is defined by
\begin{align}
\hat{\slashb{D}}=\gamma_\mu \hat{D}_\mu=
\frac{1}{2a}\sum_{\mu=1}^{4}\gamma_\mu (\hat{U}_{\mu}-\hat{U}_{-\mu}).   \label{DiracOp}
\end{align}
The eigenvalue equation of the Dirac operator can be expressed as 
\begin{eqnarray}
\hat{\slashb{D}}|n\rangle =i\lambda_n|n \rangle, \ \ \ \ \lambda_n \in {\bf R} \label{DiracEigeneq}
\end{eqnarray}
because of the anti-hermiticity of the Dirac operator. 
These Dirac eigenstates $|n\rangle$ have the completeness of $\sum_n |n\rangle\langle n|=1$. 
Due to $\{\hat{\slashb{D}},\gamma_5\}=0$, 
the chiral partner 
$\gamma_5|n\rangle$ is also an eigenstate with the eigenvalue ($-i\lambda_n$). 
Known as the Banks-Casher relation \cite{BanksCasher}, 
the low-lying Dirac modes have the dominant contribution 
to the chiral condensate $\langle \bar{\psi}\psi \rangle$ 
and thus these modes are essential modes for chiral symmetry breaking. 
\section{Dirac spectrum representation of Polyakov loop fluctuations}
In this section, 
we shortly show the derivation of the Dirac spectrum representations of the Polyakov loop fluctuations 
and then discuss the relation between confinement and chiral symmetry breaking in QCD. 
The detailed derivation is shown in Ref. \cite{DRSS, SDI, DSI}. 

We consider a temporally odd-number lattice 
with odd-number temporal size $N_\tau$. 
On such a lattice, we introduce a functional trace $I$ defined as 
\begin{eqnarray}
I={\rm Tr}_{c,\gamma} (\hat{U}_4\hat{\slashb{D}}^{N_\tau-1}),  \label{I}
\end{eqnarray}
where
${\rm Tr}_{c,\gamma}\equiv \sum_s {\rm tr}_c
{\rm tr}_\gamma$, and  ${\rm tr}_\gamma$ is taken over spinor indexes. 
Substituting the definition of the Dirac operator (\ref{DiracOp}), 
the functional trace $I$ is expressed as a sum of products of odd-number link-variable operators 
because $N_\tau$ is odd number. 
Note that most of the terms in the expansion of $I$ exactly vanish 
because one cannot make a closed loop by using odd-number link-variable operators on a square lattice. 
Thus there is only contribution from the closed path due to the temporal periodicity, 
that is the Polyakov loop $L$. 
In this way, the functional trace $I$ can be expressed as 
\begin{align}
I
=\frac{12V}{(2a)^{N_\tau-1}}L. \label{I1}
\end{align}
On the other hand, using the completeness of the Dirac mode, the functional trace is expressed as
\begin{align}
I
=\sum_n\langle n|\hat{U}_4\slashb{\hat{D}}^{N_\tau-1}|n\rangle
=i^{N_\tau-1}\sum_n\lambda_n^{N_\tau-1}\langle n|\hat{U}_4| n \rangle.  \label{I2}
\end{align}
Therefore, 
we derive the Dirac spectrum representation of the Polyakov loop: 
\begin{eqnarray}
L=\frac{(2ai)^{N_\tau-1}}{12V}
\sum_n\lambda_n^{N_\tau-1}\langle n|\hat{U}_4| n \rangle.
  \label{RelOrig}
\end{eqnarray}

Since the identity (\ref{RelOrig}) is exactly satisfied for each gauge-configuration,  
the $Z_3$ rotated Polyakov loop $\tilde{L}$ can be also expressed by the Dirac modes as 
\begin{eqnarray}
\tilde{L}=\frac{(2ai)^{N_\tau-1}}{12V}
\sum_n\lambda_n^{N_\tau-1}\mathrm{e}^{2\pi ki/3}\langle n|\hat{U}_4| n \rangle,
  \label{LtildeDirac}
\end{eqnarray}
where $k$ is defined in Eq. (\ref{Polyakov}). 
Correspondingly, 
we can express $L_L \equiv {\rm Re}(\tilde{L})$, $L_T \equiv {\rm Im}(\tilde{L})$, and $|L|$ 
by the Dirac modes. 
From these expressions and Eqs. (\ref{ChiA})-(\ref{RART}), 
one can derive analytical relations connecting the Polyakov loop fluctuations and the Dirac modes. 
In particular, we focus on the Dirac spectrum representation of $R_A$:
\begin{align}
R_A=
\frac{
\left\langle\left|\sum_n\lambda_n^{N_\tau-1}\langle n|\hat{U}_4| n \rangle\right|^2\right\rangle
-
\left\langle\left|\sum_n\lambda_n^{N_\tau-1}\langle n|\hat{U}_4| n \rangle\right|\right\rangle^2
}{
\left\langle\left(\sum_n\lambda_n^{N_\tau-1}{\rm Re}\left(\mathrm{e}^{2\pi ki/3}\langle n|\hat{U}_4| n \rangle\right)\right)^2\right\rangle
-
\left\langle\sum_n\lambda_n^{N_\tau-1}{\rm Re}\left(\mathrm{e}^{2\pi ki/3}\langle n|\hat{U}_4| n \rangle\right)\right\rangle^2
}. \label{RADirac}
\end{align}
Note that the series of analytical relations is exactly satisfied in both full QCD and quenched QCD. 

As mentioned above, the ratio $R_A$ is a sensitive probe for the quark deconfinement 
and the Dirac modes are strongly related to chiral symmtery breaking. 
Since Eq. (\ref{RADirac}) is an analytical relation connecting the ratio $R_A$ and the Dirac modes, 
we can extract from it 
the information of the relation between confinement and chiral symmetry breaking in QCD. 
The damping factors $\lambda_n^{N_\tau-1}$ in sums over all the Dirac modes 
appearing in Eq. (\ref{RADirac}) play important roles. 
Because these damping factors are negligibly small with small eigenvalues $|\lambda_n|\simeq0$, 
the contribution from the low-lying Dirac modes to the ratio $R_A$ is strongly suppressed 
while these modes are responsible for saturating the chiral condensate. 
In other words, 
important modes for chiral symmetry breaking are not important for chiral symmetry breaking. 
This analytical discussion is consistent with other numerical studies of the Dirac mode expansion \cite{GIS}, 
which show that confinement properties such as the Polyakov loop and 
the confining linear force defined from the Wilson loop 
are almost unchanged after removal of the low-lying Dirac modes. 

Not only the damping factor $\lambda_n^{N_\tau-1}$, 
but also the Dirac-mode matrix element $\langle n|\hat{U}_4| n \rangle$ should be taken into account. 
If the matrix element increases stronger than the damping factor as $\sim 1/\lambda_n^{N_\tau-1}$ in the small-$|\lambda|$ region, 
the above analytical observation is not correct. 
However, the matrix element $\langle n|\hat{U}_4| n \rangle$ 
does not change our analytical expectation qualitatively. 
The detailed numerical results of the nontrivial behavior of the matrix element 
$\langle n|\hat{U}_4| n \rangle$ are shown in Ref. \cite{DSI}.

\section{Numerical analysis}
In this section, we show the numerical results based on the analytical relation (\ref{RADirac}) 
to quantitatively confirm the above analytical discussion 
that the low-lying Dirac modes have negligible contribution to the ratio $R_A$. 

\begin{figure}
\begin{center}
\includegraphics[width=10cm]{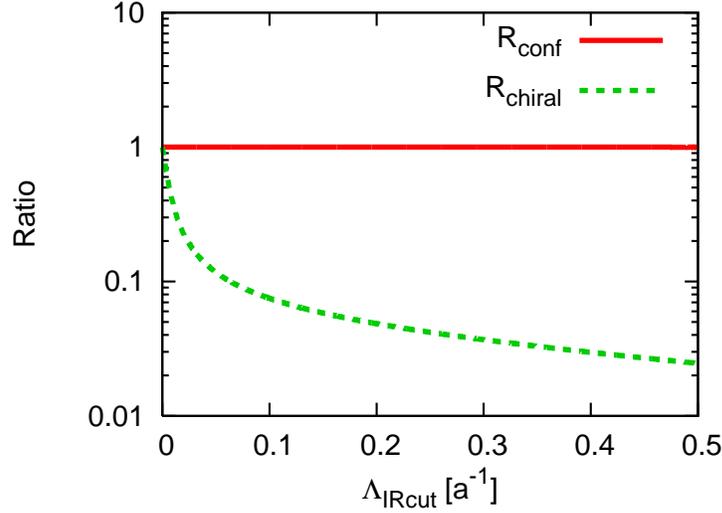}
\caption{
The numerical results for the $R_{\rm chiral}$ and $R_{\rm conf}$ 
plotted against the infrared cutoff $\Lambda$ in lattice units at $\beta=5.6$ on $10^3\times 5$ lattice 
taken from Ref. \cite{DRSS}. 
The quark mass of $m=5 \ {\rm MeV}$ is taken for calculation of the ratio $R_{\rm chiral}$. 
}
\label{ratio_conf}
\end{center}
\end{figure}

Since the Dirac eigenvalue $\lambda_n$ 
and the Dirac-mode matrix element $\langle n|\hat{U}_4| n \rangle$ are gauge-invariant 
and can be calculated by solving the Dirac eigenvalue equation Eq. (\ref{DiracEigeneq}), 
we can calculate the contribution from each Dirac mode. 
We introduce the infrared cutoff $\Lambda$ and define the $\Lambda$-dependent Polyakov loops, 
\begin{align}
&|L|_{\Lambda}=\frac{(2a)^{N_\tau-1}}{12V}
\left|\sum_{|\lambda_n|>\Lambda}\lambda_n^{N_\tau-1}\langle n|\hat{U}_4| n \rangle\right|, \label{LabsDiraccutKS}
\end{align}
for the modulus, and
\begin{align}
&(L_L)_{\Lambda}=C_{\tau}
\sum_{|\lambda_n|>\Lambda}
\lambda_n^{N_\tau-1}{\rm Re}\left(\mathrm{e}^{2\pi ki/3}\langle n|\hat{U}_4| n \rangle\right),
\label{LLDiraccutKS} \\
&(L_T)_{\Lambda}=C_{\tau}
\sum_{|\lambda_n|>\Lambda}
\lambda_n^{N_\tau-1}{\rm Im}\left(\mathrm{e}^{2\pi ki/3}\langle n|\hat{U}_4| n \rangle \right).
\label{LTDiraccutKS}
\end{align}
for the real and the imaginary part, respectively, 
with $C_{\tau}=(2ai)^{N_\tau-1}/{12V}$. 
Then, the $\Lambda$-dependent Polyakov loop susceptibilities, 
$(\chi_A)_{\Lambda}$, $(\chi_L)_{\Lambda}$, $(\chi_T)_{\Lambda}$, are defined 
using the corresponding Polyakov loops, $|L|_{\Lambda}$, $(L_L)_{\Lambda}$ or $(L_T)_{\Lambda}$,
and their ratios are also defined as 
\begin{align}
(R_A)_{\Lambda}=\frac{(\chi_A)_{\Lambda}}{(\chi_L)_{\Lambda}}, \ \ \ \ \ \ 
(R_T)_{\Lambda}=\frac{(\chi_T)_{\Lambda}}{(\chi_L)_{\Lambda}} \label{RTcut}.
\end{align}

For comparison, we introduce the cutoff-dependent chiral condensate $\langle \bar{\psi}\psi\rangle_{\Lambda}$ as 
\begin{align}
\langle \bar{\psi}\psi \rangle_{\Lambda}
=-\frac{1}{V}\sum_{|\lambda_n|\geq\Lambda}\frac{2m}{\lambda_n^2+m^2},
\end{align}
where $m$ is the current quark mass. 
In order to see the contribution from the low-lying Dirac modes 
to the ratio $R_A$ and chiral condensate, 
we introduce two quantities $R_{\rm conf}$ and $R_{\rm chiral}$ as 
\begin{align}
R_{\rm conf}= \frac{(R_A)_{\Lambda}}{R_A}, 
\ \ \ \ \ \ \ R_{\rm chiral}= \frac{\langle \bar{\psi}\psi\rangle_{\Lambda}}
{\langle \bar{\psi}\psi\rangle}. \label{Ratios}
\end{align}

The ratios $R_{\rm conf}$ and $R_{\rm chiral}$ 
are calculated in the SU(3) lattice QCD at the quenched level. 
In our calculation, the standard plaquette action is used 
on $10^3\times 5$ lattice with two different $\beta\equiv\frac{2N_{\rm c}}{g^2}$: 
$\beta=5.6$ for the confinement phase 
and $\beta=6.0$ for the deconfinement phase. 
For each $\beta$, 
20 gauge-configurations are taken every 500 sweeps after 
the thermalization of 5000 sweeps. 

In Fig. \ref{ratio_conf}, 
the numerical results for $R_{\rm conf}$ and $R_{\rm chiral}$ in the confinement phase 
are shown with various values of the infrared cutoff $\Lambda$. 
We take the quark mass $m=5$ MeV for the chiral condensate. 
From Fig. \ref{ratio_conf}, 
one can confirm that the quantity $R_{\rm chiral}$ 
is largely reduced by removal of the low-lying Dirac modes 
and then the low-lying Dirac modes are important for chiral symmetry breaking. 
Nevertheless, the quantity $R_{\rm conf}$, that is the ratio $R_A$, 
is almost unchanged by removing the low-lying Dirac modes 
even with large cutoff $\Lambda\simeq 0.5 \ {\rm GeV}$. 
The results in the deconfinement phase show the similar behavior with the confinement phase. 
In this way, we have numerically confirmed that 
the low-lying Dirac modes have little contribution to the ratio $R_A$. 

\section{Summary}
We have derived the Dirac spectrum representations of 
the Polyakov loop and its fluctuations on the temporally odd-number lattice. 
Based on the analytical relations, 
it is semi-analytically and numerically shown that 
the low-lying Dirac modes have negligible contribution to the Polyakov loop fluctuations, 
which are good probes for the quark deconfinement 
although these modes have dominant contribution to the chiral condensate. 
That is, the important modes for chiral symmetry breaking are not important for confinement. 
Our results suggest no direct one-to-one corresponding between 
confinement and chiral symmetry breaking in QCD.

\end{document}